\documentclass[11pt]{article}
\usepackage{tikz}
\usetikzlibrary{patterns}
\usetikzlibrary{arrows,shapes}
\usetikzlibrary{trees}
\usetikzlibrary{matrix,arrows} 				% For commutative diagram
\usetikzlibrary{positioning}				% For "above of=" commands
\usetikzlibrary{calc,through}				% For coordinates
\usetikzlibrary{decorations.pathreplacing}  % For curly braces
\usepackage{pgffor}							% For repeating patterns
\usetikzlibrary{decorations.pathmorphing}	% For Feynman Diagrams
\usetikzlibrary{decorations.markings}
\tikzset{
	vector/.style={decorate, decoration={snake}, draw},
	provector/.style={decorate, decoration={snake,amplitude=2.5pt}, draw},
	antivector/.style={decorate, decoration={snake,amplitude=-2.5pt}, draw},
	fermion/.style={draw=black, postaction={decorate},
		decoration={markings,mark=at position .55 with {\arrow[draw=black]{>}}}},
	fermionbar/.style={draw=black, postaction={decorate},
		decoration={markings,mark=at position .55 with {\arrow[draw=black]{<}}}},
	fermionnoarrow/.style={draw=black},
	gluon/.style={decorate, draw=black,
		decoration={coil,amplitude=4pt, segment length=5pt}},
	scalarbar/.style={dashed,draw=black, postaction={decorate},
		decoration={markings,mark=at position .55 with {\arrow[draw=black]{<}}}},
	scalarnoarrow/.style={dashed,draw=black},
	electron/.style={draw=black, postaction={decorate},
		decoration={markings,mark=at position .55 with {\arrow[draw=black]{>}}}},
	bigvector/.style={decorate, decoration={snake,amplitude=4pt}, draw},
	source/.style={draw=black, postaction={decorate}},
	binary/.style={draw=black,double distance=0.08cm, postaction={decorate}},
	bgrav/.style={dashed,draw=black, line width=1.5pt, postaction={decorate}},
	rgrav/.style={decorate, decoration={snake}, draw},
	scalar/.style={dashed,draw=black, postaction={decorate}},
	bscalar/.style={dotted,draw=black, postaction={decorate}},
	rscalar/.style={dashed,draw=black, postaction={decorate},
		decoration={markings,mark=at position .55 with {\arrow[draw=black]{>}}}},
	bphoton/.style={dashed,draw=black,double distance=0.03cm, line width=1.5pt, postaction={decorate}},
	rphoton/.style={decorate,double distance=0.02cm, decoration={snake}, draw},
	momen/.style={draw=black, postaction={decorate}, decoration={markings,mark=at position 1 with {\arrow[draw=black]{>}}}},
	cut/.style={postaction={draw,decorate,decoration={border,angle=90,amplitude=0.15cm,segment length=1mm}}}
}
\tikzstyle{block} = [draw, rectangle, 
minimum height=3em, minimum width=6em]

\usepackage[margin=1in]{geometry}
\usepackage{graphicx}
\usepackage{caption}
\usepackage{subcaption}
\usepackage{mathtools}
\usepackage{verbatimbox}
\usepackage{url}
\usepackage[square,numbers]{natbib}
\usepackage{verbatim}
\usepackage{indentfirst}
\usepackage{physics}
\usepackage{amsmath}
\usepackage{authblk}
\usepackage{physics}
\usepackage{amsmath, amsfonts, amssymb}
\usepackage{xcolor}
\usepackage[]{hyperref}
\hypersetup{
	colorlinks=true,
	linkcolor={black},
	citecolor={cyan!80!black},
	filecolor={red!60!green},      
	urlcolor={magenta!60!black},
}
\usepackage{mdframed}
\usepackage{subfiles}
\usepackage{subcaption}
\usepackage{float}

\title{EFT approach to General Relativity: Correction to EIH Lagrangian due to electromagnetic charge}

\author[]{Raj Patil\thanks{patil.raj@students.iiserpune.ac.in}}
\affil[]{Indian Institute of Science Education and Research Pune, Pune 411008, India}
\date{\today}

\begin{document}
\maketitle

\begin{abstract}
We extend the Non-Relativistic formulation of General Relativity (NRGR) given in \cite{Goldberger:2004jt} to incorporate the effects of electromagnetic charge of the constituents of the binary. We incorporate the photon field in NRGR by giving the field decomposition and power counting rules. Using these, we develop the Feynman rules to describe photon and graviton interactions with point particle worldline. We then find the corrections to the Einstein-Infeld-Hoffmann (EIH) Lagrangian \cite{Einstein:1938yz} due to the presence of the photon field and electromagnetic charge of the constituents of the binary.
\end{abstract}

%\tableofcontents

\section{Introduction}

The recent direct detection of gravitational waves by the LIGO collaboration \cite{Abbott:2016blz} has proven to be a useful probe in unraveling the mysteries of cosmic origins, equations of state of neutron stars, and have proved to be a test for the theory of General Relativity. The powerful techniques of Quantum Field Theories and General Relativity have allowed us to put forward bold claims about the binary mergers, which can then be tested by LIGO. One such technique is the tower of effective field theories given by Goldberger and Rothstein in \cite{Goldberger:2004jt}. Effective field theories can be used to describe many physical scenarios, but this was specifically made to describe the binary in the limit of small velocities of the constituents, hence the name Non-Relativistic General Relativity (NRGR).

In this analysis, we aim to extend NRGR formalism to include the effects of the electromagnetic charge of the constituents of the binary on its dynamics. A similar analysis was done previously by adding a scalar field to the NRGR formalism in \cite{Kuntz:2019zef}. Here we add a photon field to the NRGR formalism and study its interactions with the other fields. Our analysis leads to a correction to the Einstein-Infeld-Hoffmann (EIH) Lagrangian at 1PN which agrees with the previous analysis \cite{Gorbatenko2005} in the corresponding limit of charge larger than the mass of the constituents of the binary.

The outline of this paper is as follows. In section 2, we give a brief description of the NRGR, list out its assumptions, and describe the scheme used for the separation of scales. In section 3, we describe the power counting scheme and list the relevant Feynman rules. Then we detail the calculation of the effective Lagrangian up to 1PN in section 4. In section 5, we conclude the analysis with some future directions.

\section{NRGR}\label{sec_NRGR}
At first glance, it seems unnecessary to use the machinery of quantum field theories for classical calculations, but it provides us with two advantages; namely, it encapsulates divergences into standard renormalization procedures and contains the power counting techniques which allow us to calculate the order at which a given term in the perturbation series first contributes to a given physical observable. In addition to this, the separation of scales in the problem allows us to set up the calculation in a more systematic fashion, as described below.
 
In this formalism, one assumes the binary to be made up of two point particles and the internal structure of the constituents is encoded in the effective theory attached to the action of point particles. The total action governing the dynamics of the binary is given by
\begin{equation}\label{eq_S_total}
	S_{\text{Total}} = 2M_{pl}^2 \int d^4x~ \sqrt{-g} R + \int d^4x~\sqrt{-g} \Big[\frac{1}{4 \mu_0}g^{\mu\alpha}g^{\nu\beta}F_{\mu\nu}F_{\alpha\beta}\Big] + S_{\text{pp}}^{(a)}
\end{equation}
where the first term is the Einstein-Hilbert action that governs the dynamics of the underlying spacetime denoted by $S_{\text{EH}}$ with $M_{pl}$ being the Planck's length. The second term governs the dynamics of the electromagnetic field and is denoted by $S_{\text{EM}}$ where, $F_{\mu\nu} = \nabla_\mu A_\nu - \nabla_\nu A_\mu$ is the electromagnetic stress energy tensor minimally coupled to the background, $A_\mu$ is the corresponding photon field, and $\mu_0$ is the magnetic permeability of vacuum. The term $S_{\text{pp}}^{(a)}$ is the action for each constituent of the binary (approximated as point particles) and contains their interaction with the electromagnetic field as well as the underlying spacetime. It is given by
\begin{equation}\label{eq_S_pp}
	S_{\text{pp}}^{(a)} = \int ds~ \Big[-m_{(a)}\sqrt{- g_{\mu\nu}U^\mu_{(a)} U^\nu_{(a)}} +  q_{(a)} U^\mu_{(a)} A_\mu \Big] + S^{(a)}_{\text{EFT}}
\end{equation}
where $m_{(a)}$, $q_{(a)}$ and $U^\mu_{(a)}$ are the mass, electromagnetic charge and the four velocity of the $a^{\text{th}}$ point particle respectively. The effective Lagrangian $S^{(a)}_{\text{EFT}}$ contains the effective operators and Wilson coefficients that encode the internal structure of the constituents of the binary. We will ignore this term in this analysis as its leading contribution to conservative sector for a spherically symmetric object starts at 2.5PN for acceleration induced multipole moments \cite{Galley:2010es}, 3PN for multipole moments describing the tidal effects due to external electromagnetic waves \cite{Goldberger:2005cd} and 5PN for multipole moments describing the tidal effects due to external gravitational waves \cite{Goldberger:2004jt}.

Here we adopt the convention $G_N = 1/(32 \pi M^2_{pl})$, signature of the metric is mostly minus and follow the natural units and set $\hbar = c = 1$. We use Einstein’s summation notation to simplify the calculation. The Greek indices run over 0, 1, 2 and 3 and three vectors are denoted by bold Latin indices.

\subsection{Length scales}
In the bound state of two point particles, we have three length scales, namely the length scale associated with the compact object $R_s$ (Schwarzschild radius), the radius of the orbit $r$ and the wavelength of the emitted gravitational wave $\lambda$. We assume the velocities of the particles to be small as compared to the velocity of light, then we have a hierarchy of length scales
\begin{equation}
\lambda \gg r \gg R_s.
\end{equation}
The leading order behavior of the system is governed by an $1/r$ Newtonian or Coulomb interaction. Then the velocity of the point particles ($v$) relates the size of the objects to the radius of the binary and is also the ratio of the emitted wavelength and the radius of the binary given by
\begin{equation}
v^2 \approx \frac{R_s}{r} ~~~~~\text{and}~~~~~ v \approx \frac{r}{\lambda} 
\end{equation}
respectively. So it makes $v/c$ a good candidate for the expansion parameter. Another parameter we have is
\begin{equation}
  \frac{\mu_o q_{(1)} q_{(2)}}{G_N m_{(1)} m_{(2)}} = \epsilon.
\end{equation}
In this report we consider three cases: first, $\epsilon\ll 1$ which describes the gravitational force being dominant with corrections from electromagnetic force. Here we have $\epsilon$ as another expansion parameter with $v/c$. Second, $\epsilon = 1$ which describes the system with both forces of equal magnitude and thus we don't have any other expansion parameter. Third, $1/\epsilon\ll 1$ which describes the electromagnetic force being dominant with corrections from gravitational force. Here we have $1/\epsilon$ as another expansion parameter with $v/c$.

Another small parameter at hand is $\hbar/L$, where $L$ is the angular momentum of the binary. We use $v/c$ and $\epsilon$ (or $1/\epsilon$) as the primary expansion parameters where the n$^{\text{th}}$ order in $v^2/c^2$ is called as the nPN and use $\hbar/L$ to suppress the quantum effects (loop diagrams) in theory. Some of these quantum effects are studied in \cite{Goldberger:2019sya}.

\subsection{Field decomposition}
Here we assume that the particles are far from each other and hence propagate on a flat background $\eta_{\mu\nu}$. So we can decompose the metric as  
\begin{equation}
g_{\mu\nu} = \eta_{\mu\nu} + \frac{h_{\mu\nu}}{M_{pl}}
\end{equation}
where the gravitational effects of binary are propagated by the gravitons $h_{\mu\nu}$. The effective action $S_{eff}$ that describes the dynamics of binary is obtained after integrating over the gravitons and photons given by 
\begin{equation}\label{eq_S_eff_NRGR}
e^{iS_{eff}[x_{pp}^{(a)}] } = \int D[h_{\mu\nu}]~D[A_\alpha]~ e^{iS_{EH}[g_{\mu\nu}] + iS_{EM}[A_\alpha,g_{\mu\nu}] + iS_{int}[x_{pp}^{(a)},g_{\mu\nu},A_\alpha] }
\end{equation}
where $x_{pp}^{(a)}$ is the position of the $a^{\text{th}}$ point particle in the binary, and the integration is over the gauge fixed fields. As we are only interested in the long-distance physics at the scales of $\lambda$, we first decompose the fields in short distance modes - potential gravitons ($H_{\mu\nu}$) and long-distance modes - radiation gravitons ($\bar{h}_{\mu\nu}$) as 
\begin{equation}\label{eq_field_decompositionh}
h_{\mu\nu} = H_{\mu\nu} + \bar{h}_{\mu\nu}~.
\end{equation}
Similar decomposition is done for the photon field to obtain long distance modes - radiation photons ($\bar{A}_\mu$) and short distance modes - potential photons ($\textbf{A}_\mu$) as 
\begin{equation}\label{eq_field_decompositiona}
A_\mu = \textbf{A}_\mu + \bar{A}_\mu~.
\end{equation} 
This decomposition is done according to \cite{Goldberger:2004jt} by identifying the modes by their scaling with respect to $v$ and $r$. The long distance modes represent the emitted on-shell fields and its momentum scales as
\begin{equation}
\big(k_0, {\textbf{k}}\big) \approx \big(\frac{v}{r}, \frac{v}{r}\big)
\end{equation}
because $k_0$ is the frequency of the excitation and should scale like the inverse of period of the binary given by $r/v$ and so does $\textbf{k}$ because these are on-shell. 
Whereas, the short distance modes are off-shell and its momentum scales as
\begin{equation}\label{eq_k_bfk_scaling_potential_grav}
\big(k_0, \textbf{k}\big) \approx \big(\frac{v}{r}, \frac{1}{r}\big)
\end{equation}
because the wave vector $\textbf{k}$ has the dimensions of the inverse length and the only parameter at this scale is the distance between the particles in the binary given by $1/r$ and $k_0$ is the frequency of the excitation and should scale like the inverse of period of the binary given by $r/v$.

Due to the decomposition given in equation (\ref{eq_field_decompositionh}) and (\ref{eq_field_decompositiona}), we can break the integration over the perturbation shown in equation (\ref{eq_S_eff_NRGR}) in two steps. First, to describe the effective dynamics of the point particle, we construct an EFT with a systematic non-relativistic expansion in $v/c$ by integrating out the short distance modes. This effective action contains the effective binding potential for point particles and its interaction with the radiation gravitons and photons, which is responsible for the gravitational and electromagnetic wave emission. Next, integrating out the long distance radiation modes give us the dynamics of the binary and power emitted by it in the form of radiation. Only the first step is relevant to our aim and thus is described thoroughly in the next section.

%The internal structure of the point particles $R_s$ is described by the coefficients in $S_{int}$ given by equation (\ref{eq_S_eff_internal_structure}) and the internal structure of the binary $r$ is described by the potential gravitons.

\section{Conservative Dynamics}\label{Conservative_Dynamics}
For calculating the binding potential of the binary, we only consider the dynamics of the potential modes, its interactions with point particles and ignore the radiation modes from now on in this paper.
This describes the conservative dynamics of the binary given by the effective action $W_{eff}[x_{p}]$,   
\begin{equation}\label{eq_W_eff_BP}
e^{iW_{eff}[x_{pp}^{(a)}]} = \int D[H_{\mu\nu}]~ D[\textbf{A}_\alpha]~ e^{iS_{EH}[H_{\mu\nu}] + iS_{EM}[H_{\mu\nu}, \textbf{A}_\alpha] + iS_{int}[x_{pp}^{(a)},H_{\mu\nu},\textbf{A}_\alpha] }~~.
\end{equation}
%where $S_{EH}[H_{\mu\nu}]$ describes the dynamics and self-interaction of the potential graviton and the $S_{int}$ given by equation (\ref{eq_S_eff_internal_structure}) describes the dynamics of point particle and its interaction with potential gravitons. \
Note that the back-reaction on the point particles due to its interaction with a single potential graviton is of the order of $|\textbf{k}|/|\textbf{p}| \approx 1/L \ll 1$, so, the point particles are treated as background non-dynamical sources for computing the dynamics in the conservative sector \cite{Goldberger:2004jt}.

In this section, we describe the technique of power counting and give all the essential Feynman rules for propagators and vertices. Using these Feynman rules, we calculate the leading order correction to the binding potential at 1PN in the next section.%is unknown to us and we write a EFT for this. 

\subsection{Propagators}
We consider the equation (\ref{eq_S_total}), where the quadratic term gives us the propagator and the higher order terms give us the nonlinear interaction vertices. All the calculations from now on are carried out in the harmonic gauge for gravitons \cite{Donoghue:1994dn,Kol:2010si} and Feynman–’t Hooft gauge for photons. Due to small velocities of the point particles, the momentum space propagators for gravitons and photons can be expanded in series as
\begin{equation}\label{eq_corr_prop_BP}
\frac{1}{k^2}\approx\frac{-1}{\textbf{k}^2}\Big[1+\frac{k_0^2}{\textbf{k}^2}+\frac{k_0^4}{\textbf{k}^4}+ \cdot\cdot\cdot \Big]~.
\end{equation}
Due to the different scaling of $k_0$ and $\textbf{k}$ for the potential gravitons given in equation (\ref{eq_k_bfk_scaling_potential_grav}), it is convenient to perform a partial Fourier transform of the fields. Thus the momentum space Feynman rule for the propagator of the potential graviton is obtained using the leading order term in the above equation and is given by 
%\begin{mdframed}
\vspace{0.3cm}\\
\begin{minipage}{.5\textwidth}
	\centering
	\begin{tikzpicture}[line width=1 pt, scale=1]
	%\node at (-2,0) {~};
	\draw[bgrav] (0:2)--(0,0);
	\draw[momen] (0.8,-0.3)--(1.2,-0.3);
	\node at (1,-0.5) {p};
	\node at (0:3) {$H_{\alpha\beta}(t_2,\textbf{p})$};
	\node at (180:1) {$H_{\mu\nu}(t_1,\textbf{p})$};
	\end{tikzpicture}
\end{minipage}%
\begin{minipage}{.5\textwidth}
	\noindent
	\begin{flalign}
	\equiv ~~ \frac{-iP_{\mu\nu\alpha\beta}}{\textbf{p}^2}\delta(t_1-t_2)~~~~~~~~~~~~~~~~~~~~~~~~~~~~~
	\end{flalign}
\end{minipage}
%\end{mdframed}
\vspace{0.3cm}\\
where $P^{\mu\nu\alpha\beta} =  \frac{1}{2}\big(\eta^{\mu\alpha}\eta^{\nu\beta} + \eta^{\mu\beta}\eta^{\nu\alpha} - \eta^{\mu\nu}\eta^{\alpha\beta}\big)$ and higher orders of $k_0^2/\textbf{k}^2$ are taken into account by the propagator correction given by equation (\ref{eq_FR_porp_corr_bgrav}). 
Similarly at leading order, the Feynman rule for the propagator of the potential photons in momentum space is given by
%\begin{mdframed}
\vspace{0.3cm}\\
\begin{minipage}{.5\textwidth}
	\centering
	\begin{tikzpicture}[line width=1 pt, scale=1]
	\draw[bphoton] (0:2)--(0,0);
	\draw[momen] (0.8,-0.3)--(1.2,-0.3);
	\node at (1,-0.5) {p};
	\node at (0:3) {$\textbf{A}_{\nu}(t_2,\textbf{p})$};
	\node at (180:1) {$\textbf{A}_{\mu}(t_1,\textbf{p})$};
	\end{tikzpicture}
\end{minipage}%
\begin{minipage}{.5\textwidth}
	\noindent
	\begin{flalign}
	\equiv ~~ \frac{-i\eta_{\mu\nu}}{\textbf{p}^2}\delta(t_1-t_2)~~~~~~~~~~~~~~~~~~~~~~~~~~~~~
	\end{flalign}
\end{minipage}
%\end{mdframed}
\vspace{0.3cm}\\
and higher orders of $k_0^2/\textbf{k}^2$ are taken into account by the propagator correction given by equation (\ref{eq_FR_porp_corr_bphoton}). 
\subsection{Power counting}
To arrange the terms in the Lagrangian in increasing order of expansion parameter, we need to have a manifest power counting in $v$. For this, we need to know the scaling behaviour of the field and its derivatives. At the leading order, the propagator for potential modes scales as $v/r^2$
and so the potential gravitons and photons scale as 
\begin{equation}
H_{\mu\nu}(x^0,\textbf{k}) \approx \sqrt{v}r^2~~~~\text{and}~~~~\textbf{A}_{\alpha}(x^0,\textbf{k}) \approx \sqrt{v}r^2.	
\end{equation}
The scaling of its derivatives can be identified by using the Fourier transform of the potential modes. The temporal derivative scales as $v/r$, whereas the spacial derivative scales as $1/r$.
The velocity of the particles in orbit depends on the strength of the dominant force.  In the case of $\epsilon\ll 1$, the virial theorem for a gravitationally
bound system, where the motion of the constituents have typical non-relativistic velocities, give $Gm/r \approx v^2$. Thus, the gravitational coupling constant has the scaling behaviour dictated by $M_{pl}/m \approx 1/\sqrt{Lv}$ and the electromagnetic coupling constant by  $\mu_0 q^2 \approx \epsilon Lv$. Similarly, in the case of $1/\epsilon\ll 1$, using the virial theorem, the electromagnetic coupling constant has the scaling behaviour dictated by $\mu_0 q^2 \approx Lv$ and the gravitational coupling constant by $M_{pl}/m \approx 1/\sqrt{\epsilon Lv}$. Using the above scaling, we can figure out the scaling of all possible terms in the effective action. For a Feynman diagram, the propagators do not scale, but the vertices do, and their scaling is decided by the corresponding term in the Lagrangian.

\subsection{Vertices}
The action given in equation (\ref{eq_W_eff_BP}) has three kinds of interaction vertices: first, which are generated by the correction to the propagator due to the expansion given in equation (\ref{eq_corr_prop_BP}), second, which are generated by the cubic and higher order terms in $S_{\text{EH}}$ and $S_{\text{EM}}$ and third, due to the $S_{\text{pp}}^{(a)}$.

For the first kind of vertex, the scaling of the corrections to the propagator given in equation (\ref{eq_corr_prop_BP}) is given by $k_0^2/\textbf{k}^2  \approx v^2$. We define the momentum space Feynman rule for a two-point graviton interaction vertex as,
%\begin{mdframed}
\vspace{0.3cm}\\
\begin{minipage}{.5\textwidth}
	\centering
	\begin{tikzpicture}[line width=1 pt, scale=1]
	\draw[bgrav] (0:1.5)--(0,0);
	\node at (0:2.5) {$H_{\mu\nu}(t_2,\textbf{p}_2)$};
	\draw[bgrav] (180:1.5)--(0,0);
	\node at (180:2.5) {$H_{\mu\nu}(t_1,\textbf{p}_1)$};
	
	\draw[momen] (-1,-0.3)--(-0.6,-0.3);
	\node at (-0.8,-0.5) {p$_1$};
	
	\draw[momen] (1,-0.3)--(0.6,-0.3);
	\node at (0.9,-0.5) {p$_2$};
	\node at (0,0.5) {$v^2$};
	
	\draw[fill=black] (0,0) circle (.2cm);
	\draw[fill=white] (0,0) circle (.19cm);
	\clip (0,0) circle (.2cm);
	\draw[line width=1 pt] (-135:1) -- (45:1);
	\draw[line width=1 pt] (135:1) -- (-45:1);
	\end{tikzpicture}
\end{minipage}%
\begin{minipage}{.5\textwidth}
	\noindent
	\begin{flalign}\label{eq_FR_porp_corr_bgrav}
	\equiv ~~ \int dt ~ \delta^3\left(\sum_{k=1}^{2} \textbf{p}_k\right)~\Big(\partial_{t_1} \partial_{t_2}\Big)~~~~~~~~~~~
	\end{flalign}
\end{minipage}
%\end{mdframed}
\vspace{0.3cm}\\
to reproduce the propagator at higher orders in $\textbf{v}_a^2$. Similarly, the higher order correction to a two-point photon propagator is given by
%\begin{mdframed}
\vspace{0.3cm}\\
\begin{minipage}{.5\textwidth}
	\centering
	\begin{tikzpicture}[line width=1 pt, scale=1]
	\draw[bphoton] (0:1.5)--(0,0);
	\node at (0:2.5) {$\textbf{A}_{\mu}(t_2,\textbf{p}_2)$};
	\draw[bphoton] (180:1.5)--(0,0);
	\node at (180:2.5) {$\textbf{A}_{\mu}(t_1,\textbf{p}_1)$};
	
	\draw[momen] (-1,-0.3)--(-0.6,-0.3);
	\node at (-0.8,-0.5) {p$_1$};
	
	\draw[momen] (1,-0.3)--(0.6,-0.3);
	\node at (0.9,-0.5) {p$_2$};
	\node at (0,0.5) {$v^2$};
	
	\draw[fill=black] (0,0) circle (.2cm);
	\draw[fill=white] (0,0) circle (.19cm);
	\clip (0,0) circle (.2cm);
	\draw[line width=1 pt] (-135:1) -- (45:1);
	\draw[line width=1 pt] (135:1) -- (-45:1);
	\end{tikzpicture}
\end{minipage}%
\begin{minipage}{.5\textwidth}
	\noindent
	\begin{flalign}\label{eq_FR_porp_corr_bphoton}
	\equiv ~~ \int dt ~ \delta^3\left(\sum_{k=1}^{2} \textbf{p}_k\right)~\Big(\partial_{t_1} \partial_{t_2}\Big)~.~~~~~~~~~~
	\end{flalign}
\end{minipage}
%\end{mdframed}
\vspace{0.3cm}\\
Higher orders of correction can be obtained by attaching multiple vertices, each producing a factor of $k_0^2/\textbf{k}^2$ in the propagator.

For the second kind of vertex, the cubic term of the $S_{EH}[H_{\mu\nu}]$ scales as $v^2/\sqrt{L}$ which can be represented by the three-point graviton vertex given by the Feynman rule
%\begin{mdframed}
\vspace{0.3cm}\\
\begin{minipage}{.5\textwidth}
	\centering
	\begin{tikzpicture}[line width=1 pt, scale=1]
	\begin{scope}[shift={(12.5,0)}]
	\node at (1,0) {$v^2/\sqrt{L}$};
	\draw[bgrav] (90:1)--(0,0);
	\draw[bgrav] (-30:1.2)--(0,0);	
	\draw[bgrav] (-150:1.2)--(0,0);
	\end{scope}
	\end{tikzpicture}
\end{minipage}%
\begin{minipage}{.5\textwidth}
	\noindent	
	\begin{equation}\label{eq_vertex_bgrav_1}
	\text{given in equation (A.33) of \cite{Cardoso:2008gn}.}
	\end{equation}
\end{minipage}
%\end{mdframed}
\vspace{0.3cm}\\
All the higher order terms in $H_{\mu\nu}$ will have an extra factor of $M_{pl}$ in the denominator and thus scale with the higher power of $v$. In addition to this, we also have a vertex coming from the kinetic term of the photon field which scales as $v^2/\sqrt{L}$.
This corresponds to a vertex given by the Feynman rule
%\begin{mdframed}
\vspace{0.3cm}\\
\begin{minipage}{.5\textwidth}
	\centering
	\begin{tikzpicture}[line width=1 pt, scale=1.3]
	\draw[bphoton] (90:1)--(0,0);
	\draw[bgrav] (-30:1)--(0,0);
	\draw[bphoton] (210:1)--(0,0);	
	\node at (1,0) {$v^2/\sqrt{L}$};
	\end{tikzpicture}
\end{minipage}%
\begin{minipage}{.5\textwidth}
	\noindent
	\begin{equation}
	\text{given in the appendix A of \cite{BjerrumBohr:2002sx}.}
	\end{equation}
\end{minipage}
%\end{mdframed}
\vspace{0.3cm}\\
We here aim to calculate the binding potential up to order $Lv^2$ and all the diagrams having vertices contributing to higher order in $v	$ are ignored.

For the third kind of vertex, $S^{(a)}_{\text{pp}}$ action given in (\ref{eq_S_pp}) can expanded as
\begin{align}\label{eq_BP_expan_pp}
  \sum_{a=1,2}\int ds ~&\Bigg[-m_{(a)}\sqrt{ -g_{\mu\nu}\frac{dx^\mu}{ds}\frac{dx^\nu}{ds}} +  q_{(a)} \frac{dx^\mu}{ds} A_\mu \Bigg] ~d^4x~\delta^4\big(x^\mu-x_{(a)}^\mu(s)\big) =\nonumber\\ &-\sum_{a=1,2} m_{(a)}\int dt~ \Big[ 1 - \frac{1}{2}\textbf{v}_{(a)}^2+\frac{1}{2}\frac{H_{00}}{M_{pl}}+\frac{H_{0i}}{M_{pl}}\textbf{v}_{(a)}^i+\frac{1}{2}\frac{H_{ij}}{M_{pl}}\textbf{v}_{(a)}^i\textbf{v}_{(a)}^j-\frac{1}{8}\textbf{v}_{(a)}^4\nonumber\\
    &~~~~~~~~~~~~~~~~~~~~~~~~~~~~~~~~~~~~~~~~~~~~~~~~~~~~~~~~~~~~+\frac{1}{4}\frac{H_{00}}{M_{pl}}\textbf{v}_{(a)}^2+\frac{1}{8}\left(\frac{H_{00}}{M_{pl}}\right)^2+\cdot\cdot\cdot\Big]\nonumber\\
& -\sum_{a=1,2} q_{(a)}\int dt ~\Big[ \textbf{A}_0\sqrt{\mu_0} - \textbf{v}_{(a)}^i \textbf{A}_i\sqrt{\mu_0} \Big]~.
\end{align}
The scaling of terms independent of the graviton and photon field representing the kinetic energy and corrections to it is given by 
\begin{equation}
\int dt~ m~\textbf{v}^2 \approx \frac{r}{v} m v^2 \approx L~\text{;} ~~~~~ \int dt~ m~\textbf{v}^4 \approx \frac{r}{v} m v^4 \approx Lv^2~~~~~ \text{and so on.}
\end{equation}
As these terms do not contain $H_{\mu\nu}$, they do not contribute in Feynman diagrams and are carried over to the $W_{eff}[x_{p}]$. The Feynman rules and the scaling of the linear term in graviton field, coupling to the point particle worldline are given by
\vspace{0.3cm}\\
\begin{minipage}{.4\textwidth}
	\centering
	\begin{tikzpicture}[line width=1 pt, scale=1]
	\draw[source] (180:1)--(0,0);
	\draw[source] (0:1)--(0,0);
	\draw[bgrav] (-90:1)--(0,0);
	\node at (-70:1.4) {$H_{00}(t,\textbf{p})$};
	
	\draw[momen] (0.3,-0.3)--(0.3,-0.7);
	\node at (0.5,-0.5) {p};
	\node at (1.5,0) {$\textbf{x}_{(a)}(t)$};
	\node at (0,0.2) {$\sqrt{L}$};
	\end{tikzpicture}
\end{minipage}%
\begin{minipage}{.6\textwidth}
	\noindent
	\begin{flalign}\label{eq_vertex_bgrav_2}
	\equiv ~~ -\frac{im_{(a)}}{2M_{pl}} \int dt~ e^{i\textbf{p}\cdot \textbf{x}_{(a)}(t)}~;~~~~~~~~~~~~~~~~~
	\end{flalign}
\end{minipage}
%\end{mdframed}
\vspace{0.3cm}\\
%\begin{mdframed}
\vspace{0.3cm}\\
\begin{minipage}{.4\textwidth}
	\centering
	\begin{tikzpicture}[line width=1 pt, scale=1]
	\draw[source] (180:1)--(0,0);
	\draw[source] (0:1)--(0,0);
	\draw[bgrav] (-90:1)--(0,0);
	\node at (-70:1.4) {$H_{0i}(t,\textbf{p})$};
	
	\draw[momen] (0.3,-0.3)--(0.3,-0.7);
	\node at (0.5,-0.5) {p};
	\node at (1.5,0) {$\textbf{x}_{(a)}(t)$};
	\node at (0,0.2) {$\sqrt{L}v$};
	\end{tikzpicture}
\end{minipage}%
\begin{minipage}{.6\textwidth}
	\noindent
	\begin{flalign}\label{eq_vertex_bgrav_3}
	\equiv ~~ -\frac{im_{(a)}}{M_{pl}} \int dt~ \textbf{v}_{(a)}^i e^{i\textbf{p}\cdot \textbf{x}_{(a)}(t)}~~~~\text{and}~~~~~~~~~~~~~~
	\end{flalign}
\end{minipage}
%\end{mdframed}
\vspace{0.3cm}\\
%\begin{mdframed}
\vspace{0.3cm}\\
\begin{minipage}{.4\textwidth}
	\centering
        \begin{tikzpicture}[line width=1 pt, scale=1]
	\begin{scope}[shift={(0,0)}]
          \draw[source] (180:1)--(0,0);
	  \draw[source] (0:1)--(0,0);
	  \draw[bgrav] (-90:1)--(0,0);
	  \node at (-70:1.4) {$H_{ij}(t,\textbf{p})$};
	  
	  \draw[momen] (0.3,-0.3)--(0.3,-0.7);
	  \node at (0.5,-0.5) {p};
	  \node at (1.5,0) {$\textbf{x}_{(a)}(t)$};
	  \node at (0,0.2) {$\sqrt{L}v^2$};
	\end{scope}
        \begin{scope}[shift={(1.8,-0.7)}]
          \node at (0,0) {$+$};
	\end{scope}
        \begin{scope}[shift={(3.5,0)}]
          \draw[source] (180:1)--(0,0);
	  \draw[source] (0:1)--(0,0);
	  \draw[bgrav] (-90:1)--(0,0);
	  \node at (-70:1.4) {$H_{00}(t,\textbf{p})$};
	  
	  \draw[momen] (0.3,-0.3)--(0.3,-0.7);
	  \node at (0.5,-0.5) {p};
	  \node at (1.5,0) {$\textbf{x}_{(a)}(t)$};
	  \node at (0,0.2) {$\sqrt{L}v^2$};
	\end{scope}
        \end{tikzpicture}
\end{minipage}%
\begin{minipage}{.6\textwidth}
	\noindent
	\begin{flalign}\label{eq_vertex_bgrav_4}
	\equiv ~~ -\Bigg[\frac{im_{(a)}}{2M_{pl}} \int dt~ \textbf{v}_{(a)}^i \textbf{v}_{(a)}^j e^{i\textbf{p}\cdot \textbf{x}_{(a)}(t)}  + \frac{im_{(a)}}{4M_{pl}} \int dt~ \textbf{v}_{(a)}^2 e^{i\textbf{p}\cdot \textbf{x}_{(a)}(t)}\Bigg]~.~~~~~~~~~~~~~~~~~
	\end{flalign}
\end{minipage}
%\end{mdframed}
\vspace{0.3cm}\\
where, the solid straight line is used to represent the world line point-particles which do not propagate because they do not carry any momentum. Similarly, the linear term in photon field coupling to the point particle worldline has the following Feynman rules and scaling,
\vspace{0.3cm}\\
\begin{minipage}{.4\textwidth}
	\centering
	\begin{tikzpicture}[line width=1 pt, scale=1]
	\draw[source] (180:1)--(0,0);
	\draw[source] (0:1)--(0,0);
	\draw[bphoton] (-90:1)--(0,0);
	\node at (-70:1.4) {$\textbf{A}_{0}(t,\textbf{p})$};
	
	\draw[momen] (0.3,-0.3)--(0.3,-0.7);
	\node at (0.5,-0.5) {p};
	\node at (1.5,0) {$\textbf{x}_{(a)}(t)$};
	\node at (0,0.2) {$\sqrt{L}$};
	\end{tikzpicture}
\end{minipage}%
\begin{minipage}{.6\textwidth}
	\noindent
	\begin{flalign}
	\equiv ~~ -iq_{(a)}\sqrt{\mu_0} \int dt~ e^{i\textbf{p}\cdot \textbf{x}_{(a)}(t)}~~~~~\text{and}~~~~~~~~~~~~~
	\end{flalign}
\end{minipage}
%\end{mdframed}
\vspace{0.3cm}\\
%\begin{mdframed}
\vspace{0.3cm}\\
\begin{minipage}{.4\textwidth}
	\centering
	\begin{tikzpicture}[line width=1 pt, scale=1]
	\draw[source] (180:1)--(0,0);
	\draw[source] (0:1)--(0,0);
	\draw[bphoton] (-90:1)--(0,0);
	\node at (-70:1.4) {$\textbf{A}_{i}(t,\textbf{p})$};
	
	\draw[momen] (0.3,-0.3)--(0.3,-0.7);
	\node at (0.5,-0.5) {p};
	\node at (1.5,0) {$\textbf{x}_{(a)}(t)$};
	\node at (0,0.2) {$\sqrt{L}v^2$};
	\end{tikzpicture}
\end{minipage}%
\begin{minipage}{.6\textwidth}
	\noindent
	\begin{flalign}
	\equiv ~~ -iq_{(a)}\sqrt{\mu_0} \int dt~ \textbf{v}^i_{(a)} e^{i\textbf{p}\cdot \textbf{x}_a(t)}~~.~~~~~~~~~~~~~~~~
	\end{flalign}
\end{minipage}
%\end{mdframed}
\vspace{0.3cm}\\

The Feynman rule and scaling of the leading quadratic term in graviton field, coupling with the point particle worldline is given by
\vspace{0.3cm}\\
\begin{minipage}{.5\textwidth}
	\centering
	\begin{tikzpicture}[line width=1 pt, scale=1]
	\draw[source] (180:1)--(0,0);
	\draw[source] (0:1)--(0,0);
	\draw[bgrav] (-50:1.5)--(0,0);
	\draw[bgrav] (-130:1.5)--(0,0);
	\draw[momen] (0.7,-0.5)--(1,-0.8);
	\node at (1,-0.5) {p};
	\draw[momen] (-0.7,-0.5)--(-1,-0.8);
	\node at (-1,-0.5) {q};
	\node at (1.5,0) {$\textbf{x}_{(a)}(t)$};
	\node at (0,0.2) {$v^2$};
	\node at (-45:2.1) {$H_{00}(t,\textbf{p})$};
	\node at (-115:1.6) {$H_{00}(t,\textbf{q})$};
	\end{tikzpicture}
\end{minipage}%
\begin{minipage}{.5\textwidth}
	\noindent
	\begin{flalign}\label{eq_vertex_bgrav_5}
	\equiv ~~ -\frac{im_a}{8M_{pl}} \int dt~  e^{i\textbf{p}\cdot \textbf{x}_{(a)}(t)}e^{i\textbf{q}\cdot \textbf{x}_{(a)}(t)}~.~~~~~~~~~~~~~~~~~
	\end{flalign}
\end{minipage}
%\end{mdframed}
\vspace{0.3cm}\\
The terms that contribute to a higher order of $v$ are ignored here. 

In all the above diagrams, the vertices have to be contracted with the propagator having the corresponding free indices. For example, in the case of gravitons, vertex with $\textbf{v}^i$ or $\textbf{v}^i\textbf{v}^j$ has to be contracted with $P_{i0\_\_}$ or $P_{ij\_\_}$ respectively. If a vertex does not have a free index then it should be contracted with $P_{00\_\_}$. Similarly in the case of photons, vertex with $\textbf{v}^i$  has to be contracted with $\eta_{i\_}$ and ff a vertex does not have a free index then it should be contracted with $\eta_{0\_}$. The order of $\epsilon$ that each vertex contributes to is not shown in the above given Feynman rules. This is different for the case of $\epsilon\ll 1$ and $1/\epsilon \ll 1$, which could be easily seen by the scaling of $M_{pl}$ and $\mu_0$ with $\epsilon$ in each vertex for each case. Because the Feynman rules are given in momentum space, all the undetermined 3-momentas of photons and gravitons have to be integrated over.

\section{Calculating the binding potential}
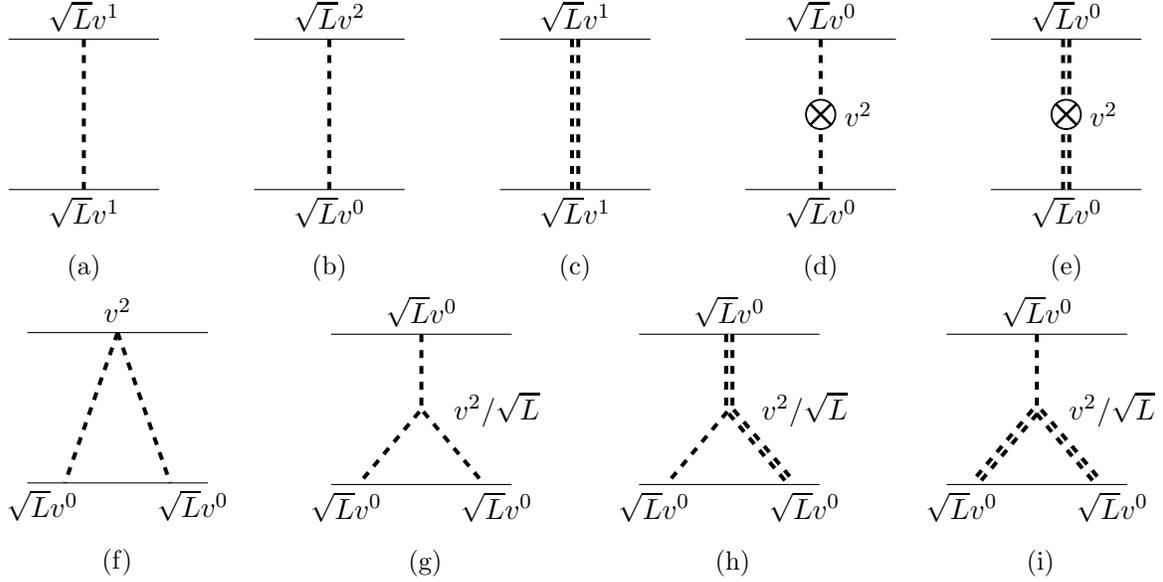
\begin{figure}[t]
\centering
\begin{subfigure}{0.19\textwidth}
\centering
\begin{tikzpicture}%[line width=1 pt, scale=0.8]
\draw[source] (-1,1)--(1,1);
\draw[source] (-1,-1)--(1,-1);
\draw[bgrav] (0,1)--(0,-1);
%\node at (1.5,1) {$\textbf{x}_1(t)$};
%\node at (1.5,-1) {$\textbf{x}_2(t)$};
\node at (0,1.3) {$\sqrt{L}v^1$};
\node at (0,-1.3) {$\sqrt{L}v^1$};
\end{tikzpicture}	
\caption{}
\label{fig_lv2_a}
\end{subfigure}
\begin{subfigure}{0.19\textwidth}
\centering
\begin{tikzpicture}%[line width=1 pt, scale=0.8]
\draw[source] (-1,1)--(1,1);
\draw[source] (-1,-1)--(1,-1);
\draw[bgrav] (0,1)--(0,-1);
%\node at (1.5,1) {$\textbf{x}_1(t)$};
%\node at (1.5,-1) {$\textbf{x}_2(t)$};
\node at (0,1.3) {$\sqrt{L}v^2$};
\node at (0,-1.3) {$\sqrt{L}v^0$};
\end{tikzpicture}	
\caption{}
\label{fig_lv2_b}
\end{subfigure}
\begin{subfigure}{0.19\textwidth}
\centering
\begin{tikzpicture}%[line width=1 pt, scale=0.8]
\draw[source] (-1,1)--(1,1);
\draw[source] (-1,-1)--(1,-1);
\draw[bphoton] (0,1)--(0,-1);
%\node at (1.5,1) {$\textbf{x}_1(t)$};
%\node at (1.5,-1) {$\textbf{x}_2(t)$};
\node at (0,1.3) {$\sqrt{L}v^1$};
\node at (0,-1.3) {$\sqrt{L}v^1$};
\end{tikzpicture}	
\caption{}
\label{fig_lv2_c}
\end{subfigure}
\begin{subfigure}{0.19\textwidth}
\centering
\begin{tikzpicture}%[line width=1 pt, scale=0.8]
\draw[source] (-1,1)--(1,1);
\draw[source] (-1,-1)--(1,-1);
\draw[bgrav] (0,1)--(0,-1);
%\node at (1.5,1) {$\textbf{x}_1(t)$};
%\node at (1.5,-1) {$\textbf{x}_2(t)$};
\node at (0,1.3) {$\sqrt{L}v^0$};
\node at (0,-1.3) {$\sqrt{L}v^0$};
\node at (0.5,0) {$v^2$};
\draw[fill=black] (0,0) circle (.2cm);
\draw[fill=white] (0,0) circle (.19cm);
\clip (0,0) circle (.2cm);
\draw[line width=1 pt] (-135:1) -- (45:1);
\draw[line width=1 pt] (135:1) -- (-45:1);
\end{tikzpicture}	
\caption{}
\label{fig_lv2_d}
\end{subfigure}
\begin{subfigure}{0.19\textwidth}
\centering
\begin{tikzpicture}%[line width=1 pt, scale=0.8]
\draw[source] (-1,1)--(1,1);
\draw[source] (-1,-1)--(1,-1);
\draw[bphoton] (0,1)--(0,-1);
%\node at (1.5,1) {$\textbf{x}_1(t)$};
%\node at (1.5,-1) {$\textbf{x}_2(t)$};
\node at (0,1.3) {$\sqrt{L}v^0$};
\node at (0,-1.3) {$\sqrt{L}v^0$};
\node at (0.5,0) {$v^2$};
\draw[fill=black] (0,0) circle (.2cm);
\draw[fill=white] (0,0) circle (.19cm);
\clip (0,0) circle (.2cm);
\draw[line width=1 pt] (-135:1) -- (45:1);
\draw[line width=1 pt] (135:1) -- (-45:1);
\end{tikzpicture}	
\caption{}
\label{fig_lv2_e}
\end{subfigure}
\newline
\begin{subfigure}{0.24\textwidth}
\centering
\begin{tikzpicture}%[line width=1 pt, scale=0.8]
\node at (0,1.3) {$v^2$};
\node at (1,-1.3) {$\sqrt{L}v^0$};
\node at (-1,-1.3) {$\sqrt{L}v^0$};
\draw[source] (-1.2,1)--(1.2,1);
\draw[bgrav] (0,1)--(0.7,-1);
\draw[bgrav] (0,1)--(-0.7,-1);	
\draw[source] (-1.2,-1)--(1.2,-1);
\end{tikzpicture}	
\caption{}
\label{fig_lv2_f}
\end{subfigure}
\begin{subfigure}{0.24\textwidth}
\centering
\begin{tikzpicture}%[line width=1 pt, scale=0.8]
\node at (0,1.3) {$\sqrt{L}v^0$};
\node at (1,-1.3) {$\sqrt{L}v^0$};
\node at (-1,-1.3) {$\sqrt{L}v^0$};
\node at (1,0) {$v^2/\sqrt{L}$};
\draw[source] (-1.2,1)--(1.2,1);
\draw[bgrav] (90:1)--(0,0);
\draw[bgrav] (-50:1.2)--(0,0);	
\draw[bgrav] (-130:1.2)--(0,0);
\draw[source] (-1.2,-1)--(1.2,-1);
\end{tikzpicture}	
\caption{}
\label{fig_lv2_g}
\end{subfigure}
\begin{subfigure}{0.24\textwidth}
\centering
\begin{tikzpicture}%[line width=1 pt, scale=0.8]
\node at (0,1.3) {$\sqrt{L}v^0$};
\node at (1,-1.3) {$\sqrt{L}v^0$};
\node at (-1,-1.3) {$\sqrt{L}v^0$};
\node at (1,0) {$v^2/\sqrt{L}$};
\draw[source] (-1.2,1)--(1.2,1);
\draw[bphoton] (90:1)--(0,0);
\draw[bphoton] (-50:1.2)--(0,0);	
\draw[bgrav] (-130:1.2)--(0,0);
\draw[source] (-1.2,-1)--(1.2,-1);
\end{tikzpicture}	
\caption{}
\label{fig_lv2_h}
\end{subfigure}
\begin{subfigure}{0.24\textwidth}
\centering
\begin{tikzpicture}%[line width=1 pt, scale=0.8]
\node at (0,1.3) {$\sqrt{L}v^0$};
\node at (1,-1.3) {$\sqrt{L}v^0$};
\node at (-1,-1.3) {$\sqrt{L}v^0$};
\node at (1,0) {$v^2/\sqrt{L}$};
\draw[source] (-1.2,1)--(1.2,1);
\draw[bgrav] (90:1)--(0,0);
\draw[bphoton] (-50:1.2)--(0,0);	
\draw[bphoton] (-130:1.2)--(0,0);
\draw[source] (-1.2,-1)--(1.2,-1);
\end{tikzpicture}	
\caption{}
\label{fig_lv2_i}
\end{subfigure}
\caption{The diagrams that contribute to order $Lv^2$.}
\label{fig_lv2}
\end{figure}

In this section, we explicitly calculate the first few terms of the effective Lagrangian for the action given in equation (\ref{eq_W_eff_BP}). We have obtained this action by integrating out potential gravitons and photons and ignoring the radiation ones. This corresponds to the diagrams having the property that they are connected after removing the worldlines of the particles, internal lines correspond to potential gravitons and/or photons, and we do not have external on-shell potential modes \cite{Porto:2016pyg,Levi:2018nxp}. We ignore all the diagrams that contain graviton and/or photon loops because adding a loop adds a factor of $v^4/L$ to the diagrams and suppresses it by a huge factor of $1/L$. We also ignore all the diagrams with graviton and/or photon self-energy terms, which can be made zero by the techniques of dimensional regularization.\footnote{A few self-energy divergent contributions are still present in diagrams \ref{fig_lv2_g}, \ref{fig_lv2_h} and \ref{fig_lv2_i}, but could be handled easily by textbook regularization techniques.}

\begin{figure}[t]
\centering
\begin{subfigure}{0.49\textwidth}
\centering
\begin{tikzpicture}%[line width=1 pt, scale=0.8]
\draw[source] (-1,1)--(1,1);
\draw[source] (-1,-1)--(1,-1);
\draw[bgrav] (0,1)--(0,-1);
%\node at (1.5,1) {$\textbf{x}_1(t)$};
%\node at (1.5,-1) {$\textbf{x}_2(t)$};
\node at (0,1.3) {$\sqrt{L}v^0$};
\node at (0,-1.3) {$\sqrt{L}v^0$};
\end{tikzpicture}	
\caption{}
\label{fig_lv0_a}
\end{subfigure}
\begin{subfigure}{0.49\textwidth}
\centering\begin{tikzpicture}%[line width=1 pt, scale=0.8]
\draw[source] (-1,1)--(1,1);
\draw[source] (-1,-1)--(1,-1);
\draw[bphoton] (0,1)--(0,-1);
%\node at (1.5,1) {$\textbf{x}_1(t)$};
%\node at (1.5,-1) {$\textbf{x}_2(t)$};
\node at (0,1.3) {$\sqrt{L}v^0$};
\node at (0,-1.3) {$\sqrt{L}v^0$};
\end{tikzpicture}	
\caption{}
\label{fig_lv0_b}
\end{subfigure}
\caption{The diagrams that contribute to order $Lv^0$.}
\label{fig_lv0}
\end{figure}
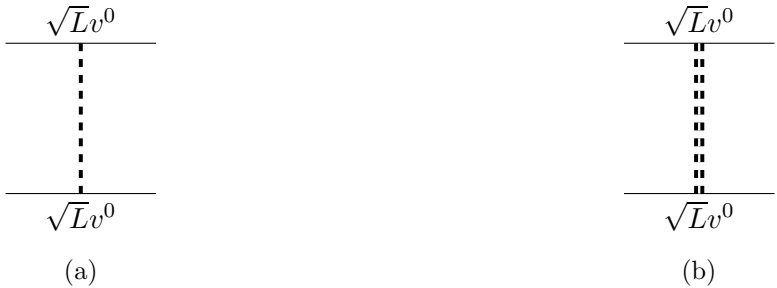

The diagrams contributing at order $Lv^0$ (for arbitrary order of $\epsilon$) are given in figure \ref{fig_lv0}. The equation (\ref{app_eq_coloumb_int}) in the appendix is used to compute the integrals in corresponding to the diagrams. The corresponding Lagrangian is given by
\begin{equation}\label{eq_newton_coulomb}
L_{(\text{Lv}^0)} = \sum_{a=1,2} \frac{1}{2} m_{(a)} \textbf{v}_{(a)}^2 + \frac{G_{N} m_{(1)}m_{(2)}}{|\textbf{r}|}- \frac{\mu_0}{4\pi} \frac{q_{(1)}q_{(2)}}{|\textbf{r}|}
\end{equation}
where $\textbf{r} = \textbf{x}_{(1)}-\textbf{x}_{(2)}$. In the above Lagrangian, the first term represents the kinetic energy of the particle, the second is the Newtonian gravitational potential and the third is the Coulomb potential.

The diagrams contributing at order $Lv^2$ (for arbitrary order of $\epsilon$) are given in figure \ref{fig_lv2}. 
The diagrams \ref{fig_lv2_a}, \ref{fig_lv2_b}, \ref{fig_lv2_c} and \ref{fig_lv2_f}  are straightforward to compute using equation (\ref{app_eq_coloumb_int}) in the appendix. For the diagram \ref{fig_lv2_d} and \ref{fig_lv2_e} we use equation (\ref{app_eq_int_idk}) in the appendix and for the diagram \ref{fig_lv2_g}, \ref{fig_lv2_h} and \ref{fig_lv2_i}, we use equations (\ref{app_eq_int_1_k}) in the appendix and (A.34) in \cite{Cardoso:2008gn}. The results then correspond to the following Lagrangian
\begin{align}\label{eq_corr_EIH}
  L_{(\text{Lv}^2)} =& \sum_{a=1,2} \frac{1}{8} m_{(a)} \textbf{v}_{(a)}^4 + \frac{G_N m_{(1)} m_{(2)}}{2|\textbf{r}|}\Big[  3\big(\textbf{v}_{(1)}^2+\textbf{v}_{(2)}^2\big)-7\big(\textbf{v}_{(1)}\cdot\textbf{v}_{(2)}\big) - \frac{\big(\textbf{v}_{(1)}\cdot\textbf{r}\big)\big(\textbf{v}_{(2)}\cdot\textbf{r}\big)}{\textbf{r}^2}\Big] \nonumber\\
  &~~~~~~~~~- \frac{G_N^2 m_{(1)}m_{(2)}\big(m_{(1)}+m_{(2)}\big)}{2\textbf{r}^2} + \frac{\mu_0}{4\pi} \frac{q_{(1)}q_{(2)}}{2|\textbf{r}|} \Big[  \big(\textbf{v}_{(1)}\cdot\textbf{v}_{(2)}\big) + \frac{\big(\textbf{v}_{(1)}\cdot\textbf{r}\big)\big(\textbf{v}_{(2)}\cdot\textbf{r}\big)}{\textbf{r}^2}\Big]\nonumber \\
  &~~~~~~~~~+ \frac{G_N \mu_0}{4\pi}\Big[\frac{2q_{(1)}q_{(2)}(m_{(1)}+m_{(2)})+m_{(1)}q^2_{(2)}+m_{(2)}q^2_{(1)}}{2\textbf{r}^2}\Big].
\end{align}
%The last term in the above equation goes to zero in two cases: first, if the charges are equal and opposite and second,if the charges are opposite and the ratio $|m_1/m_2| = |q_1/q_2|$. 
The first three terms of the above Lagrangian is the Einstein-Infeld-Hoffmann Lagrangian \cite{Einstein:1938yz}. The next term is the correction to the EIH Lagrangian due to the charge of the constituents, which has also been derived using techniques of classical GR in \cite{Gorbatenko2005}. The final term corresponds to the non-linear interaction between the gravitational and electromagnetic forces. This term has been newly calculated by the techniques detailed in this paper.
%The new result of this paper is the final term in the above Lagrangian which corresponds to the to the non-linear interaction between the gravitational and electromagnetic forces.

The above results can now be analysed in the order of $\epsilon$. For the case of $\epsilon \ll 1$ the hierarchy of diagrams is given in table \ref{table_el1}. The corresponding terms in the Lagrangian could be easily recognized according to the power of $\mu_0$ as it scales like $\epsilon$. Similarly, for the case of $1/\epsilon \ll 1$ the hierarchy of diagrams is given in table \ref{table_1bel1} and the corresponding terms in the Lagrangian could be easily recognized according to the power of $G_N$ because it scales like $1/\epsilon$. For the case of $\epsilon \approx 1$, the Lagrangian at order $Lv^0$ is given in equation (\ref{eq_newton_coulomb}) and at order $Lv^2$ is given in equation (\ref{eq_corr_EIH}).

\setlength{\tabcolsep}{1em} % for the horizontal padding
{\renewcommand{\arraystretch}{1.2}% for the vertical padding
\begin{table}
    \centering
  \begin{tabular}{||c||c|c|c||} 
 \hline
  & $\epsilon^0$ & $\epsilon^1$ & $\cdots$ \\
 \hline\hline
 $v^0$ & Figure \ref{fig_lv0_a} & Figure \ref{fig_lv0_b} & \\ 
 \hline
  & Figure \ref{fig_lv2_a} & Figure \ref{fig_lv2_c} &  \\
  & Figure \ref{fig_lv2_b} & Figure \ref{fig_lv2_e} &  \\
 $v^2$ & Figure \ref{fig_lv2_d} & Figure \ref{fig_lv2_h} &  \\
  & Figure \ref{fig_lv2_f} & Figure \ref{fig_lv2_i} &  \\
  & Figure \ref{fig_lv2_g} &  &  \\
 \hline
 $\vdots$ &  &  & \\
 \hline
 \end{tabular}
  \caption{The hierarchy of diagrams at different orders of $v$ and $\epsilon$ for $\epsilon\ll 1$.}
  \label{table_el1}
\end{table}

\begin{table}
    \centering
  \begin{tabular}{||c||c|c|c|c||} 
 \hline
  & $1/\epsilon^0$ & $1/\epsilon^1$ & $1/\epsilon^2$ & $\cdots$ \\
 \hline\hline
 $v^0$ & Figure \ref{fig_lv0_b} & Figure \ref{fig_lv0_a} &  & \\ 
 \hline
  &  & Figure \ref{fig_lv2_a} &  &  \\
  & Figure \ref{fig_lv2_c} & Figure \ref{fig_lv2_b} & Figure \ref{fig_lv2_f} & \\
 $v^2$ &  & Figure \ref{fig_lv2_d} &  &\\
  & Figure \ref{fig_lv2_e} & Figure \ref{fig_lv2_h} & Figure \ref{fig_lv2_g} &  \\
  &  & Figure \ref{fig_lv2_i} & & \\
 \hline
 $\vdots$ &  &  &  &\\
 \hline
 \end{tabular}
  \caption{The hierarchy of diagrams at different orders of $v$ and $\epsilon$ for $1/\epsilon\ll 1$.}
  \label{table_1bel1}
\end{table}

\section{Conclusion}

In this paper, we have detailed the calculation of 1PN effective Lagrangian, including the effects of charge of the constituents of the binary. A similar analysis was previously done in \cite{Gorbatenko2005}, but in a limit of charge of the binary constituents much more than their mass. In such a limit, one can only derive terms that are zeroth order in $G_N$. The techniques of NRGR allow us to follow through the analysis without assuming any such limit and thus gives a general result. The result corresponds to equation (\ref{eq_corr_EIH}), where we find an additional term over the previous results of \cite{Einstein:1938yz} and \cite{Gorbatenko2005}. We then prescribe the method to systematically study the correction due to the sub dominant force in orders of $\epsilon$ and $1/\epsilon$ as given in table \ref{table_el1} and \ref{table_1bel1} respectively.

This analysis can be extended further to higher PN orders using the techniques developed in \cite{Kol:2007bc} and to include the spin of the constituents of the binary using the analysis given in \cite{Porto:2005ac}.
Here we have not discussed the effects of the radiation photons on the dynamics of the binary. Using the radiation photons, the leading order waveform and the radiated power by the binary in terms of its multipole expansion could be found in \cite{Ross:2012fc}, but the higher-order effects due to the interaction of photon and graviton are yet to be analyzed. Such corrections have to be computed analogously to the analysis done for the radiation gravitons in \cite{Goldberger:2009qd}. Also, the EFT describing the internal structure of a charged compact object contains highly non-trivial interaction terms of the gravitons, photons and the spin, and thus deserves a careful analysis.

\appendix
\section*{Acknowledgements}
I would like to express my gratitude to Dr.~Suneeta Vardarajan for the useful comments, remarks, and engagement during the analysis. I am grateful for the support provided by the INSPIRE Scholarship for Higher Education, Government of India. We also thank the referee for making important suggestions for improving the scope of this paper. Lastly, I am thankful for all the insightful discussions I had with Rahul Poddar and Palash Singh.

\appendix
\section*{Appendix}
The integrals over momentum are calculated using the following identities,
\begin{align}\label{app_eq_coloumb_int}
\int \frac{d^3k}{(2\pi)^3} \frac{e^{i\textbf{k}\cdot\textbf{r}}}{\textbf{k}^2} = \frac{1}{4\pi}\frac{1}{|\textbf{r}|}~~;
\end{align}
\begin{align}\label{app_eq_int_1_k}
\int \frac{d^3k}{(2\pi)^3} \frac{e^{i\textbf{k}\cdot\textbf{r}}}{|\textbf{k}|} = \frac{1}{2\pi^2}\frac{1}{\textbf{r}^2}~~~~\text{and}
\end{align}
\begin{align}\label{app_eq_int_idk}
\int \frac{d^{(n-1)}\textbf{q}}{(2\pi)^{(n-1)}} \frac{\textbf{q}_i\textbf{q}_j}{\textbf{q}^4}e^{i\textbf{q}\cdot\textbf{x}}= \frac{\Gamma\big(\frac{n-3}{2}\big)}{8\pi^{(\frac{n-1}{3})}}|\textbf{x}|^{3-n}\Big[\delta_{ij}+(3-n)\frac{\textbf{x}_i\textbf{x}_j}{\textbf{x}^2}\Big]~~.
\end{align}

%\newpage
\bibliographystyle{unsrt}
\bibliography{Citations}

\end{document}